\documentclass[aps,reprint,superscriptaddress,nofootinbib,floatfix,longbibliography]{revtex4-2}

\usepackage{amsmath}
\usepackage{bm}
\usepackage{amsfonts}
\usepackage{graphicx}
\usepackage{hyperref}
\usepackage{xcolor}

\enlargethispage{\baselineskip}

\hypersetup{
     colorlinks   = true,
     citecolor    = blue
}
\everymath{\displaystyle}

\newcommand{\TDLI}{\affiliation{Tsung-Dao Lee Institute (TDLI) \& School of Physics and Astronomy, Shanghai Jiao Tong University, \\ Shengrong Road 520, 201210 Shanghai, P.\ R.\ China}}

\begin{document}

\title{Consistency of the string inspired electroweak axion with cosmic birefringence}

\author{Weikang Lin}
\email{weikanglin@sjtu.edu.cn}
\TDLI

\author{Tsutomu T. Yanagida}
\email{tsutomu.tyanagida@sjtu.edu.cn }
\TDLI
\affiliation{Kavli IPMU (WPI), The University of Tokyo, Kashiwa, Chiba 277-8583, Japan}
\date{\today}

\begin{abstract}
    We revisit the constraint from the recently reported cosmic birefringence on axion-like particles with a general decay constant. Particular attention is paid to the naturalness of the model parameter space, which has been overlooked in the literature. We show that the observed cosmic birefringence is naturally explained by the electroweak axion with a string-theory inspired decay constant $F_A\simeq 10^{16}$\,GeV. 
\end{abstract}

\maketitle

\section{Introduction}

%String theory is the unique known theory as consistent quantum gravity. Finding phenomenological evidence for string theory is therefore important but has yet to be fruitful. A promising avenue is to search for axion-like particles, which commonly exist in string theory with a wide spectrum of potentially testable phenomenology in cosmology and astrophysics \cite{Arvanitaki:2009fg}. %In this work, we point out that the recently observed cosmic birefringence \cite{Minami:2020odp,Diego-Palazuelos:2022dsq,Eskilt:2022cff} is the first hint to string theory.h
%which may include low-energy supersymmetry (SUSY)and/or axion-like light pseudo scalar bosons \cite{Witten}. 

Axion-like particles commonly exist in string theory which furnishes a wide spectrum of cosmological and astrophysical phenomena \cite{Arvanitaki:2009fg}. In many string models, it is difficult to have a decay constant $F_A$ of axion drastically below $\sim 10^{16}$\,GeV \cite{Svrcek:2006yi}. On the other hand, if $F_A$ is larger than $\sim10^{16}$\,GeV, the axion quality problem becomes severe: gravitational effects would introduce a large explicit breaking of the continuous global shift symmetry required for the axion \cite{Alonso:2017avz,Giddings:1987cg}. Therefore, in string theory it is preferable to have $F_A\simeq10^{16}$\,GeV which we call in this work the ``string inspired'' decay constant.

Recently, \cite{Minami:2020odp,Diego-Palazuelos:2022dsq,Eskilt:2022cff} reported a detection of cosmic birefringence at the $\sim3\sigma$ confidence level. If not due to unaccounted-for systematic errors, this result signals some parity-violating process occurred between recombination and today  \cite{Carroll:1989vb,Harari:1992ea,Carroll:1998zi,Lue:1998mq}. While some works {explaining this tentative cosmic birefringence measurement are based on axion-like particle scenarios} \cite{Fujita:2020aqt,Fujita:2020ecn,Nakagawa:2021nme,Choi:2021aze}, their axion decay constants considered are in tension with the string inspired value. Also, their analyses are not based on a concrete particle model.

In this work, we extend the analysis of the axion-like particle scenario to include a more general value of $F_A$. We pay attention to the naturalness of the parameter space regarding the initial condition and the anomaly coefficient. We show that the most natural axion-like particle scenario that explains the reported cosmic birefringence is consistent with the electroweak (EW) axion \cite{Nomura:2000yk} with a string inspired $F_A$.

\section{The electroweak axion}
The EW axion is a Nambu-Goldstone boson whose mass is given mostly by the electroweak $SU(2)$ instanton \cite{Nomura:2000yk}. We assumed a new type of axion $A$ which couples the EW $SU(2)_L$ gauge fields $W^a_{\mu\nu}$ in the standard model with 
\begin{equation}\label{eq:EWaxionCoupling}
    \mathcal{L}\supset\frac{g_2^2}{32\pi^2} A W^a_{\mu \nu}{\widetilde{W}^{a\mu \nu}}\,,
\end{equation}
where $\widetilde{W}^{a\mu\nu}$ is the dual of $W^a_{\mu\nu}$ and $a=1,2,3$.  In this paper, we assume low-energy supersymmetry (SUSY). The potential of the EW axion $A$ is given by the $SU(2)$ instanton effects with the following form \cite{Nomura:2000yk}
\begin{equation}\label{eq:axion-potential}
     V_A=\frac{\Lambda_{A}^4}{2}\big(1-\cos(A/F_A)\big) \,,
\end{equation}
with the potential height given by
\begin{equation}\label{eq:potential-height}
\begin{split}
    \Lambda_{A}^4&\simeq2e^{-\frac{2\pi}{\alpha_2(M_{Pl})}}c\,\epsilon^{10}m_{3/2}^3M_{Pl}\\
    &\simeq c\big(\frac{\epsilon}{1/17}\big)^{10}\big(\frac{m_{3/2}}{1\,{\rm TeV}}\big)^3(1.4\times10^{-3}\,{\rm eV})^4,
\end{split}
\end{equation}
where $m_{3/2}$ is the gravitino mass, $c$ is an $O(1)$ dimensionless constant, the constant $\epsilon\simeq 1/17$ is an explicit breaking value of spurion \cite{Buchmuller:1992qc} of the Froggatt-Nielsen global symmetry (see \cite{Froggatt:1978nt})\footnote{The factor $\epsilon^{10}$ is expressed by $\sim \frac{m_c}{m_t}\frac{m_u}{m_t}\frac{m_{\mu}}{m_{\tau}}\frac{m_e}{m_{\tau}}$.} and $M_{Pl}\simeq2.4\times10^{18}$\,GeV is the reduced Planck mass. The Froggatt-Nielsen global $U(1)$ symmetry was invented to explain the observed mass hierarchies in quark and lepton mass matrices \cite{Froggatt:1978nt}. They introduce a $U(1)$ charged field $\phi$ which has a vacuum expectation value $\langle\phi\rangle=\epsilon$. Then, all masses and mixing angles are determined by the powers of $\epsilon$ and the powers are given by the corresponding charges of the Froggatt-Nielsen global $U(1)$. This mechanism is well known for its successful explanation of the mass hierarchies of quarks and leptons. Here, we have assumed the gravity mediation to estimate the SUSY-breaking soft masses of standard-model SUSY particles. A crucial point here is that the EW axion potential is determined by the electroweak gauge coupling constant $\alpha_2(M_{Pl})\simeq1/23$ at  $M_{Pl}$, for a given SUSY breaking scale $m_{3/2}$. We have taken the cut-off scale to be $M_{Pl}$. We note that the form of the potential Eq.\,\eqref{eq:axion-potential} is precise as long as the dilute gas approximation in the instanton calculus is reliable. In our case, the dilute gas approximation is reliable since the axion potential is given by the small size instantons \cite{Nomura:2000yk}.  %However, our result does not depend much on the cut-off scale, and in fact, the axion potential reduces by a factor of $10$ if we take the cut-off scale as $0.1 M_{Pl}$. %It is remarkable that the axion potential $V_A$ above is given by the electroweak instanton and the string-inspired decay constant $F_A$, besides the $O(1)$ constants. It is very interesting if this restricted model can naturally explain the observed cosmic birefringence \cite{Minami:2020odp,Diego-Palazuelos:2022dsq,Eskilt:2022cff}.

Note that SUSY is assumed in the calculation of the potential height. In the non-SUSY case, the axion potential is suppressed by a small $SU(2)$ gauge coupling constant at the Planck scale \cite{McLerran:2012mm}. However, if we assume the anomaly-free Froggatt-Nielsen discrete symmetry $Z_{10}$ we do not have the suppression factor $\epsilon^{10}$ \cite{Choi:2019jck} and recover the potential of a magnitude similar to our case \cite{McLerran:2012mm}.

The axion potential Eq.\,\eqref{eq:axion-potential} gives us the axion mass $m_A$ around the potential minimum as
\begin{equation}\label{eq:axion-mass}
     m_A=\frac{\Lambda_A^2}{\sqrt{2}F_A}\simeq\big(\frac{m'_{3/2}}{1\,{\rm TeV}}\big)^{\frac{3}{2}}\big(\frac{M_{Pl}}{F_A}\big) \times6\times10^{-34}\,{\rm eV}\,,
\end{equation}
where we take $\epsilon=1/17$ \cite{Buchmuller:1992qc} and absorb the $O(1)$ parameter $c$ into the gravitino mass to define an effective gravitino mass as $m'_{3/2} = c^{1/3} m_{3/2}$. We take $m'_{3/2} = 1-100$ TeV \cite{Ibe:2011aa,Ibe:2012hu} considering the $\mathcal{O}(1)$ ambiguity of the constant $c$. Then, for a string inspired decay constant $F_A \simeq 10^{16}$\,GeV, the EW axion has a mass $m_A\simeq 10^{-31} - 10^{-28}$\,eV. This corresponds to an interesting mass range studied in \cite{Fujita:2020aqt,Fujita:2020ecn} and we will further emphasize its importance in this work. This is a remarkable result since if the axion potential $V_A$ is not dominantly induced by the electroweak instantons but by unknown interactions, the axion mass is a completely free parameter.

The coupling of the EW axion to the electromagnetic fields is given by \cite{Choi:2021aze}
\begin{equation}\label{eq:couplingAFF}
    \mathcal{L} \supset -c_\gamma\frac{\alpha}{4\pi}\frac{A}{F_A} F_{\mu\nu}\tilde{F}^{\mu\nu}.
\end{equation}
where $\alpha\simeq1/137$ is the fine structure constant, $c_\gamma$ an anomaly coefficient, $F_{\mu\nu}$ and $\tilde{F}^{\mu\nu}$ are the Faraday tensor and its dual. For a better comparison to the work \cite{Fujita:2020aqt,Fujita:2020ecn}, we identify the EW axion-photon coupling constant as 
\begin{equation}\label{eq:coupling-constant}
    g\equiv\frac{c_\gamma\alpha}{\pi F_A}\,.
\end{equation}
Note that if $A$ only couples to the weak gauge fields as shown in Eq.\,\eqref{eq:EWaxionCoupling} as in the minimal model, the coupling Eq.\,\eqref{eq:couplingAFF} with $c_\gamma=1$ will be generated after the electroweak symmetry breaking \cite{Choi:2021aze}. However, in general $A$ can couples to the hypercharge $U(1)$ gauge field. In that case $c_\gamma$ is a free parameter. 
Now we discuss the generation of the cosmic birefringence for the above axion mass region.

\begin{figure}
    \centering
    \includegraphics[width=0.99\linewidth]{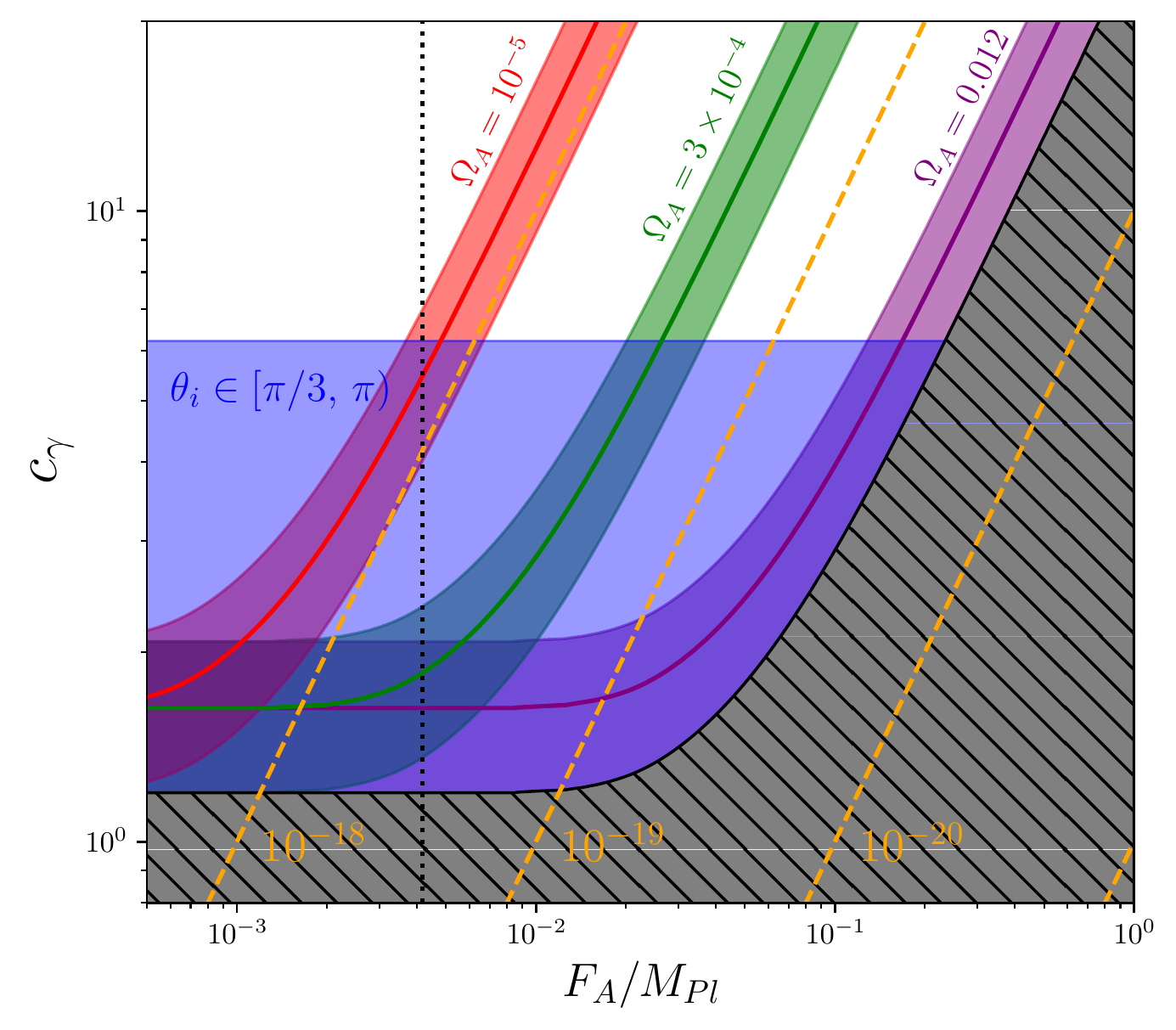}
    \caption{Constraints on the $c_\gamma$-$F_A$ parameter space. The constraint from the cosmic birefringence (1-$\sigma$) with some selected $\Omega_A$ values are shown in the purple, the green and the red bands. The gray area is excluded. The blue area shows the more natural parameter space with $\theta_i\in[\pi/3,\pi)$; see Sec. \ref{sec:naturalness} for a discussion. Orange dashed lines show constant values ($10^{-18}$, $10^{-19}$ and $10^{-20}$) of the coupling constant $g\equiv c_\gamma\alpha/\pi F_A$.  The vertical dotted line shows $F_A=10^{16}$ GeV.}
    \label{fig:parameter-space}
\end{figure}
\section{The Cosmic Birefringence}
The Cosmic Microwave Background (CMB) polarization pattern can be decomposed into an even-parity $E$ mode and an odd-parity $B$ mode. If the polarization distribution is parity invariant, the cross-correlation between the $E$ and the $B$ modes vanishes. A nonvanishing cross correlation between the two modes would then signal some parity violating physics at cosmological scales \cite{Carroll:1989vb,Harari:1992ea,Lue:1998mq}. Recently, detection of such a cross-correlation is reported in \cite{Minami:2020odp,Diego-Palazuelos:2022dsq,Eskilt:2022cff} assuming the polarization planes of CMB photons are all rotated by some angle $\beta$ in the same direction with respect to their propagation. Such a uniform rotation furnishes a $B$ mode out from the $E$ mode (with $B_{\ell m}\simeq 2\beta E_{\ell m}$ for $\beta\ll1$ for each multiple-moment) and thus a correlation between them. To date, the detection is at a $3.6\sigma$ confidence level with $\beta=0.342^\circ\,^{+0.094^\circ}_{-0.091^\circ}$ \cite{Eskilt:2022cff}. 

A promising explanation for such a uniform rotation of the CMB polarization plane is the ``cosmic birefringence'', where some dynamical scalar field (like $A$) couples to the electromagnetic fields via a Chern-Simons term \cite{Carroll:1989vb,Harari:1992ea,Carroll:1998zi,Lue:1998mq} such as Eq.\,\eqref{eq:couplingAFF}. As a CMB photon travels in an $A$-varying background, its polarization plane is rotated by an angle (the cosmic birefringence angle) given by \cite{Carroll:1989vb}
\begin{equation}\label{eq:CBangle-deltaA}
    \beta=0.21{\rm\, deg}\times c_{\gamma}\times\frac{\Delta \theta}{\pi}\,,
\end{equation}
where we have defined $\theta \equiv A/F_A$ and $\Delta \theta$ is the change of $\theta$ from recombination to today. For the considered mass range of $A$, it suffices to ignore the fluctuation of $A$ to account for the isotropic cosmic birefringence \cite{Fujita:2020aqt}.

\section{Analysis}
%Recall that we take an effective gravitino mass $m'_{3/2} = 1-100$ TeV so that the resultant EW axion mass is $m_A\simeq 10^{-31} - 10^{-28}$\,eV for a string inspired decay constant $F_A\simeq10^{16}$\,GeV.
To perform a thorough analysis, we first assume an intermediate axion mass range, i.e., $m_A\simeq 10^{-31} - 10^{-28}$\,eV but widely release the value of $F_A$ rather than fixing it to $F_A\simeq10^{16}$\,GeV. The range of $F_A$ is however finite. Recall that the axion potential Eq.\,\eqref{eq:axion-potential} is given by the electroweak instantons and the axion mass depends on $F_A$ via Eq.\,\eqref{eq:axion-mass}. The mass range $m_A\simeq 10^{-31} - 10^{-28}$\,eV and an effective gravitino mass $m'_{3/2}=1-100$\,TeV then give $F_A\sim10^{13}-10^{19}$\,GeV. We will later comment on the naturalness of this assumed mass range. Note that the analysis in \cite{Fujita:2020aqt,Fujita:2020ecn} is not applicable here as they fix the axion decay constant to $F_A=M_{Pl}$.  We find that there is some nontrivial difference between the case with $F_A=M_{Pl}$ and that with $F_A\lesssim0.01 M_{Pl}$,  and thus the axion mass is not the only phenomenologically important parameter. 

We define $\Omega_A\equiv\rho_A^0/\rho_c^0$ where $\rho_A^0$ and $\rho_c^0=3M_{Pl}^2H_0^2$ are the EW axion energy density and the critical density today. For the  mass range considered, it has been shown that the axion energy fraction today is small and bounded by $\Omega_{A}\lesssim0.006h^{-2}\simeq0.012$ \cite{Hlozek:2014lca} where $h\simeq0.7$ is the Hubble constant normalized by $100$\,km/s/Mpc. Thus, for the dynamics of the scale factor of the Universe, we can ignore the effect of $A$ and we assume a standard flat $\Lambda$CDM cosmology.

The evolution of an  axion-like field in an expanding background has been extensively studied in the literature. We refer readers to, e.g., \cite{Marsh:2010wq,Hlozek:2014lca,Fujita:2020aqt,Fujita:2020ecn} for the details of solving the dynamics of $A$. Here, we show the result most relevant to our analysis. We consider a homogeneous universe. Initially, $A$ is ``frozen'' at early times and starts to oscillate when $m_A\sim3H$ with a gradually damping amplitude. Since the amplitude of the EW axion has been sufficiently damped by today, we have $\Delta \theta\simeq \theta_{i}$ where $\theta_{i}$ is the initial value. We find that $\Omega_{A}$ is related to $\Delta \theta$ by
\begin{equation}\label{eq:Oa-theta_i-relation}
    \Omega_{A}\simeq f(\tfrac{\Delta\theta}{\pi})\times2\,\Omega_{\rm{m}}\big(\frac{F_A}{M_{Pl}}\big)^2\Delta\theta^2\,,
\end{equation}
where $\Omega_{\rm m}\simeq0.3$ \cite{Lin:2021sfs,Planck:2018vyg} is today's matter energy density fraction. Some values of $f(\tfrac{\Delta\theta}{\pi})$ are given in Table \ref{tab:fx-values}. The above approximation is good with some $<1\%$ deviation from numerical results as long as (1) $H_0\ll m_{A}\ll H_{eq}$ (which well includes the mass range considered), where $H_{eq}$ is the Hubble rate at  matter-radiation equality and (2) the initial $\theta_i$ is not fine-tuned to the potential top. A similar relation is given by Eq. (11) in \cite{Hlozek:2014lca} (also see Eq. (12) in \cite{Fujita:2020aqt}), but our Eq\,.\eqref{eq:Oa-theta_i-relation} is more general and we use an axion-like potential with a general value of $F_A$. When $\Delta\theta\lesssim0.2\pi$, our Eq\,.\eqref{eq:Oa-theta_i-relation} reduces to Eq. (11) in \cite{Hlozek:2014lca}. We however note that, since our result is based on numerical calculations, the coefficient of our Eq.\,\eqref{eq:Oa-theta_i-relation} is different from that of Eq. (11) in \cite{Hlozek:2014lca} even when $\Delta\theta\ll1$.

We note that $m_A$ is absent in both Eqs.\,\eqref{eq:CBangle-deltaA} and \eqref{eq:Oa-theta_i-relation}, but $F_A$ enters Eq.\,\eqref{eq:Oa-theta_i-relation} in a non-trivial way compared to the quadratic potential case. Thus, it is $F_A$ that becomes an important phenomenological parameter instead of $m_A$ in this intermediate axion mass range.  

\subsection{Constraints with some fixed $\Omega_A$}
With Eqs.\,\eqref{eq:CBangle-deltaA} and \eqref{eq:Oa-theta_i-relation}, we can quickly obtain the constraint from the cosmic birefringence on the $c_\gamma$-$F_A$ plane for any fixed $\Omega_A$ detailed below.

The ${\Omega_{A}M_{Pl}^2}/{F_A^2}\ll1$ case: In this case, since $f(\Delta\theta/\pi)\geq1$, we have $\Delta\theta\ll1$ inferred from \eqref{eq:Oa-theta_i-relation}. The axion-like potential reduces to a quadratic form, and from Table \ref{tab:fx-values} we have $f(\Delta\theta/\pi)\simeq1$ so that our Eq.\,\eqref{eq:Oa-theta_i-relation} reduces to Eq. (11) in \cite{Hlozek:2014lca}. As a result, the maximally allowed $\Omega_A$ corresponds to a maximally allowed $\Delta A$ from \eqref{eq:Oa-theta_i-relation} and to a minimally allowed $c_\gamma/F_A$ (and hence the coupling constant $g$) from Eq.\,\eqref{eq:CBangle-deltaA}. This is consistent with the analyses in \cite{Fujita:2020aqt,Fujita:2020ecn} where they take $F_A=M_{Pl}$ (see Figures 1 and 2 in \cite{Fujita:2020ecn} for examples). In order to explain the observed cosmic birefringence, the coupling constant is at least $g \sim3.5\times10^{-20}$\,GeV$^{-1}$, which is shown by the upper-right part of the purple band in Figure \ref{fig:parameter-space}. For the parameter space considered, the coupling constant is well below the astrophysical upper bound of $\mathcal{O}(10^{-12})$\,GeV \cite{Berg:2016ese}. We, therefore, do not show this bound in Figure \ref{fig:parameter-space}. If we take $F_A=M_{Pl}$, this means an unnaturally large $c_\gamma\gtrsim35$ is required. A $c_\gamma$ of $\mathcal{O}(1)$ may be achieved with a smaller $F_A$, but then Eq. (11) in \cite{Hlozek:2014lca} may break down and so the situation needs to be more accurately described by our Eq.\,\eqref{eq:Oa-theta_i-relation}, especially in the other extreme case as discussed below.

\begin{table}[tbp]
    \centering
\begin{ruledtabular}
    \caption{Values of $f(\tfrac{\Delta\theta}{\pi})$ defined in Eq.\,\eqref{eq:CBangle-deltaA} for some selected $\Delta\theta/\pi$. Recall that $\Delta\theta\simeq\theta_i$ in our case.}
    \label{tab:fx-values}
    \begin{tabular}{cccccccccc}
        $\Delta\theta/\pi$ & $<0.2$& $0.3$ & $0.4$ & $0.5$ & $0.6$ & $0.7$ & $0.8$ & $0.9$ & $0.99$ \\
        $f(\Delta\theta/\pi)$ & $1.0$ & $1.1$ & $1.2$ & $1.4$ & $1.6$ & $2.0$ & $2.7$ & $4.2$ & $13.7$
    \end{tabular}
\end{ruledtabular}
\end{table}

The ${\Omega_{A}M_{Pl}^2}/{F_A^2}\gg1$ case: In this case, $\Delta\theta$ is bounded ($\Delta\theta<\pi$) but the factor $f(\Delta\theta/\pi)$ becomes large and approaches to infinity as $\Delta\theta$ approaches to $\pi$. As a result, $\Delta\theta\sim\pi$ is the solution of Eq.\,\eqref{eq:Oa-theta_i-relation} once ${\Omega_{A}M_{Pl}^2}/{F_A^2}\gg1$ is satisfied. Then, the observed cosmic birefringence $\beta=0.342^\circ\,^{+0.094^\circ}_{-0.091^\circ}$ no longer corresponds to a constraint on $g$ for a fixed $\Omega_A$. Instead, it asymptotically corresponds to a constraint on the anomaly coefficient $c_\gamma=1.6\pm0.4$ inferred from Eq.\,\eqref{eq:CBangle-deltaA}, which is shown by the horizontal part of the purple band in Figure \ref{fig:parameter-space}.  Such an asymptotic constraint on $c_\gamma$ applies to all values of $\Omega_A$ as long as ${\Omega_{A}M_{Pl}^2}/{F_A^2}\gg1$.

The resultant constraint on the $c_\gamma$-$F_A$ plane from the cosmic birefringence for any fixed value of $\Omega_A$ is similar to the purple constraint in Figure \ref{fig:parameter-space}, except that it is shifted to the left for a smaller value of $\Omega_A$. In Figure \ref{fig:parameter-space}, we show in green the constraint with $\Omega_A=10^{-4}$ for a comparison. The gray parameter space is excluded by combing the observed cosmic birefringence and the bound of $\Omega_A\lesssim0.012$. Note that the minimal model with $c_{\gamma} =1$ \cite{Choi:2021aze} is consistent with the observation within a $2\sigma$ confidence level.

\subsection{The most natural parameter space}\label{sec:naturalness}
If $\Omega_A$ is not too low, it might be determined with future cosmological observations combining CMB and large-scale structures as studied, e.g., in \cite{Hlozek:2014lca}. But so far, it is only a free parameter except for that it is bounded by $\Omega_A\lesssim0.012$. Therefore, all the parameter space above the purple band in Figure \ref{fig:parameter-space} is in principle allowed. However, they are not all equally natural. First of all, $c_\gamma=1$ in the minimal model \cite{Choi:2021aze}. It is then more desirable to have a $c_\gamma$ of $\mathcal{O}(1)$. Secondly, the initial $\theta_i$ is naturally of $\mathcal{O}(1)$, then from Eq.\,\eqref{eq:CBangle-deltaA} we also have a $c_\gamma$ of $\mathcal{O}(1)$. To roughly represent these points, in Figure \ref{fig:parameter-space} we show the parameter space in blue where $\theta_i\in[\pi/3,\pi)$, which is also where $c_\gamma$ is of $\mathcal{O}(1)$. If we impose such a ``naturalness'', the parameter space with $F_A\gtrsim0.1M_{Pl}$ is excluded; see Figure \ref{fig:parameter-space}.  

Throughout the discussion, we have assumed $m_A\simeq 10^{-31} - 10^{-28}$\,eV. The importance of this mass range has been pointed out in \cite{Fujita:2020aqt,Fujita:2020ecn} that the coupling constant $g$ required to explain the cosmic birefringence can be the smallest in such a mass range. But by taking $F_A=M_{Pl}$, one is actually not able to explain the cosmic birefringence with an $\mathcal{O}(1)$ $c_\gamma$. Here, we improve the analysis to include a general $F_A$ and further emphasize the importance of such a mass range: \textit{only in this mass range with $F_A\lesssim0.1M_{Pl}$ can one explain the observed cosmic birefringence with a natural $\mathcal{O}(1)$ anomaly coefficient $c_\gamma$ without any fine-tuned initial condition.} Indeed, for a smaller axion mass, while $A$ can be a dark energy candidate, the initial field value needs to be fine-tuned to the potential top and a $c_\gamma$ of at least $\mathcal{O}(10)$ is required to explainer the observed cosmic birefringence \cite{Choi:2021aze}. On the other hand, for a larger axion mass, the oscillation started before recombination unless the initial field value is fine-tuned to the potential top. It is then difficult to have a $\Delta\theta$ of $\mathcal{O}(1)$ and so according to Eq.\,\eqref{eq:CBangle-deltaA} $c_\gamma$ is required to be much larger than $\mathcal{O}(1)$ to explain the observed cosmic birefringence. In this work, we show that in the intermediate mass range one can achieve a $c_\gamma$ of $\mathcal{O}(1)$ only with $F_A\lesssim0.1M_{Pl}$. 

So far, we have been treating $F_A$ as a free parameter. It is remarkable that, as shown by the vertical dotted line, the string inspired decay constant $F_A\simeq10^{16}$\,GeV can explain the observed cosmic birefringence while passing the requirements of naturalness. Recall that the EW axion with a string inspired $F_A$ is a rather restricted model. The EW axion potential and the axion mass are predicted by the model and it happens to fall into the mass range that allows the most natural explanation for the cosmic birefringence.  

%So far, we have been treating $F_A$ as a free parameter. However, recall that the axion mass depends on $F_A$ with Eq.\,\ref{eq:axion-mass}. Taking the most natural mass range $m_A\simeq 10^{-31} - 10^{-28}$\,eV as discussed above and an effective gravitino mass $m'_{3/2}=1-100$\,TeV, the $F_A$ is roughly required to be $10^{13}-10{19}$\,GeV. The string-inspired decay constant $F_A\simeq10^{16}$\,GeV is well within this range. Recall that the EW axion with a string inspired $F_A$ is a rather restricted model. It is remarkable that, while the EW axion potential and the axion mass are predicted by the model, it happens to fall into the mass range that allows the most natural explanation to the cosmic birefringence. 

%To further abuse the naturalness requirement of the parameter space, we can infer a lower bound\baselineskip of the energy TO BE CONTINUED???

\section{Discussion}
Note that we do not assume the presence of QCD axion. However, if one needs the QCD axion as the solution to the strong CP problem, the string-inspired $F_A$ would make the QCD axion density largely exceed the observed dark matter density \cite{Bae:2008ue}. However, this problem can be solved, e.g., by late-time entropy production \cite{Kawasaki:1995vt}. In the case of some string theory that has at least two massless axion-like bosons, $A_1$ and $A_2$, which both have anomalous couplings to the strong $SU(3)$ and weak $SU(2)$ gauge fields. In that case, we can in general define the EW axion $A$ as a linear combination of $A_1$ and $A_2$, which only couples to weak gauge fields. Then, this EW axion receives a mass only from the electroweak instantons; see details in \cite{Lin:2022khg}.

We have considered a homogeneous configuration of an axion-like field. Alternative, axionic domain walls may explain the reported cosmic birefringence  \cite{Takahashi:2020tqv,Kitajima:2022jzz}. In that case, a peculiar anisotropic cosmic birefringence is predicted \cite{Takahashi:2020tqv,Kitajima:2022jzz}. Incidentally, \cite{Jain:2022jrp} argues that it is difficult for an axion-like particle defect network to accommodate the isotropic cosmic birefringence with the non-detection of anisotropic one. 

Another possibility for the EW axion to explain the cosmic birefringence is that it behaves as a quintessence dark energy if $F_A \simeq 10^{17-18}$ GeV and the EW axion was initially around the potential top \cite{Choi:2021aze,Fukugita:1994xn,*Fukugita:1995nb,Frieman:1995pm,Nomura:2000yk,Choi:1999xn,Ibe:2018ffn}. Besides the differences in the naturalness of the value of $F_A$, the required anomaly factor $c_\gamma$ and the initial condition, the two scenarios differ in the time of onset of the cosmic birefringence. This difference may be distinguished by the cosmic birefringence tomography proposed in \cite{Sherwin:2021vgb}.

One important prediction of the EW axion with a string inspired $F_A$ and an $\mathcal{O}(1)$ $c_\gamma$ is that $\Omega_A$ is bounded from below. This can be seen from Figure \ref{fig:parameter-space} that the blue area around the vertical dotted line only allows $\Omega_A\gtrsim10^{-5}$. This might be too low compared to the current upper bound of $\Omega_A\lesssim0.012$. But, since a lower $c_\gamma$ that is closer to unity is theoretically referred, a larger $\Omega_A$ is somehow favored. For example, if we restrict $c_\gamma<2$, $\Omega_A$ would be further constrained to $\Omega_A\gtrsim10^{-4}$. Still, this is about two orders of magnitude smaller than the current upper bound. Nonetheless, if $\Omega_A$ is not too low, combining future CMB experiments and galaxy surveys may observe the effects of this EW axion on the suppression of the growth of the small-scale structures that are discussed in \cite{Hlozek:2014lca}.

\section{Conclusions}
In this work, we have shown that the electroweak axion naturally explains the observed cosmic birefringence with a string-inspired decay constant $F_A\simeq10^{16}$\,GeV. While the axion potential is generated by the electroweak $SU(2)$ instanton, the axion mass is predicted to be $m_A\simeq 10^{-31} - 10^{-28}$\,eV. We revisited the constraints from the cosmic birefringence on an axion-like field focusing on this mass range but with a general value of $F_A$. We found that only in this intermediate axion mass range ($m_A\simeq 10^{-31} - 10^{-28}$\,eV) with $F_A<0.1M_{Pl}$ can one naturally explain the observed cosmic birefringence with an $\mathcal{O}(1)$ $c_\gamma$ without any fine-tuned initial condition. Remarkably, this mass range and the bound of $F_A$ are consistent with the string-inspired  axion model. The observed cosmic birefringence may then be the first phenomenological hint of string theory.

\begin{acknowledgments}
T. T. Y. is supported in part by the China Grant for Talent Scientific Start-Up Project and by the Natural Science Foundation of China (NSFC) under grant No. 12175134 as well as by World Premier International Research Center Initiative (WPI Initiative), MEXT, Japan.
\end{acknowledgments}

%\newpage

\bibliographystyle{apsrev4-1}
\bibliography{refs}

\end{document}